\documentclass[12pt]{iopart}

\usepackage{graphicx}
\begin{document}

\title{Degeneracies in the length spectra of metric graphs}

\author{U. Gavish$^{1}$ and U. Smilansky$^2$}

\address{1. Laboratoire de Physique Th\'eorique et Hautes \'Energies, CNRS UMR 7589, Universit\'e Paris 6,
4 Place Jussieu, 75005 Paris, France\\
2. Weizmann Institute of Science, Rehovot 76100, Israel}

\begin{abstract}
The spectral theory of quantum graphs is related via an
exact trace formula with the spectrum of the lengths of periodic
orbits (cycles) on the graphs. The latter is a degenerate spectrum,
and understanding its structure (i.e., finding out how many different lengths exist for periodic orbits
with a given period and the average number of periodic orbits with the same length) is necessary for
the systematic study of spectral fluctuations using the trace formula.
This is a combinatorial problem which we solve exactly for complete (fully connected) graphs with
arbitrary number of vertices.

\end{abstract}

\maketitle

\section{Introduction}
The interest in the spectral properties of the Schr\"odinger
operator on metric graphs (known also as "quantum graphs")
increased dramatically after it was found that quantum graphs
provide an excellent paradigm for the study of spectral
fluctuations in quantum chaotic systems \cite {KS97,KS99}: The
spectral density of quantum graphs can be expressed as an {\it
exact} trace formula \cite {Roth,KS97} in terms of the spectrum
of the lengths of its periodic orbits (PO) (also called cycles)  which is analogous
to the {\it asymptotic} semi-classical trace formula \cite {Gutz}.
Moreover, the quantum graphs posses a Liouvillian analogue, which,
under some well understood conditions is ergodic. At the same
time, extensive numerical simulations and tests can be performed
with rather modest computational effort allowing detailed
comparison of the spectral statistics with the prediction of
Random Matrix Theory, and study the systematic deviation from it.
The simple finite graphs are essentially one dimensional (albeit
not simply connected) systems. They display spectral complexity
under one important condition: the lengths of the bonds must be
rationally independent.

The main tool in the theoretical discussion of the spectral
statistics of quantum graphs is the above-mentioned trace formula.
It can be written explicitly as
\begin{equation}
\label{trace1}
 d(k) = \sum_{j=1}^{\infty} \delta (k-k_j) = \frac{\mathcal{L}}{\pi}
 +\sum_n \sum_{p\in \mathcal{P}_n} A^{(n)}_p {\rm e}^{ikl_p}\ ,
\end{equation}
  where $E_j=k_j^2$ are the eigenvalues of the Schr\"odinger
operator. The $k_j$'s form the wave-number spectrum.
$\mathcal{L}$ is the total bond length, i.e., for a
graph with $B$ bonds, of lengths $L_b$, it is given by
$\mathcal{L}=\sum_{b=1}^B L_b$.  $\mathcal{P}_n$ denotes the set of PO's of
period $n$. Each periodic orbit  contributes a term which consists of a "transition amplitude"
$A^{(n)}_p$, and a unimodular factor with a phase which is
determined by the length of the corresponding $n$-bond PO
\begin{equation}
\label{lenghts}
 l_p=\sum_{b=1}^B q_b L_b \ \ \ \ ; \ \ \  q_b \in
\{0,1,2,\cdots\}\ \ {\rm and}\ \  \sum_{b=1}^B q_b = n\ .
\end{equation}
The length spectrum is highly degenerate.
Each degeneracy class contains  all orbits which traverse the same bonds the
same number of times, but not in the same order (up to cyclic
permutations) and only these (since the bond lengths $L_b$ are rationally independent).
That is, a degeneracy class of n-bond PO's
consists of orbits which have the same code ${\bf q}^{(n)}
=(q_1,\cdots,q_B)$, with $\sum_b q_b =n$.
Not every set of  nonnegative integers $\{q_b\}$ the sum of which is $n$
represents a degeneracy class - the graph connectivity and the periodicity
restrict the possible codes.
 The trace formula can be written as
\begin{equation}
\label{trace2}
 d(k) =  \frac{\mathcal{L}}{\pi}
 +\sum_n \sum _{{\bf q}^{(n)}}\ \left[ \sum_{p\in {\bf q}^{(n)}}
  A^{(n)}_p\ \right] {\rm e}^{ikl_{\bf q}^{(n)}}\ ,
\end{equation}
Where the contributions of the different orbits in the degeneracy
class ${\bf q}^{(n)}$ were lumped together to the sum in the
square brackets, all having the same phase factor.

There are two prominent examples where a detailed information
about the degeneracy classes is needed. The first example emerges in
attempts to understand the conditions under which the spectral
fluctuation of a quantum graph follow the predictions of Random
Matrix Theory. A standard tool is the computation of the spectral
autocorrelation function
\begin{equation}
\label{autocorr}
R(\xi;k) = \frac{1}{2\Delta}\int_{k-\Delta}^{k+\Delta}  \tilde d(x+\frac{\xi}{2})
\tilde d(x-\frac{\xi}{2}){\rm d }x \ ,
\end{equation}
where $ \tilde d(k) = d(k)- \frac{\mathcal{L}}{\pi}$ is the
fluctuating part of the spectral density and the domain of
integration $[k-\Delta,k+\Delta]$ is arbitrarily large.
Substituting the explicit expression (\ref {trace2}) in
(\ref{autocorr}) one sees that the autocorrelation
function depends on the squares of the individual contributions of
the degeneracy classes (the terms in square brackets in (\ref
{trace2})).

The second example is encountered in the context of "hearing the shape
of a graph", that is, in attempts to reconstruct the connectivity and
the length spectrum from the energy eigenvalue spectrum of the quantum graph \cite
{GS01}. The main tool is again the trace formula (\ref {trace2})
or rather its Fourier transform $\hat d(l) = \int {\rm d}k d(k)
\exp(ikl)$. $\hat d(l)$ is a distribution supported on the
length spectrum, with weights which can be read off from (\ref
{trace2}). The length spectrum (its composition and weights) is therefore useful
to obtain the information about $\hat d(l)$ and in turn also about the connectivity of the graph.

The leading asymptotic (for large $n$) contribution to the number of degeneracy classes
in a general connected graph was obtained by Berkolaiko \cite{BK00}.
The number of degeneracy classes and the number of PO's in each class
were obtained by Tanner \cite{Tanner} for binary graphs up to order 6.

In this paper we present an exact expression for the number
of classes for fully connected (complete) graphs of any order.

We start by defining precisely graphs, PO's, and their degeneracy classes.
We then compute the number of degeneracy classes for general fully
connected graphs, the total number of n-bond PO's, and obtain the mean degeneracy of the classes
as the ratio between the two.  Finally, we present numerical results and interpret them.

\section{Graphs, Periodic orbits, Degeneracy classes}
  A  \emph{graph,} $G,$ of order $V,$ is a set of $V$ numbered vertices, some of
which are connected by a bond (no more than one bond between two vertices, no bond connects a vertex to itself).
 The number of bonds connected to a vertex is the vertex \emph{valency.}
The \emph{connectivity matrix} of  $G$ is defined   by
\begin{eqnarray}\label{connct}
C_{ij}(G)=
 \{ \begin{array}{rr}
 1 & if ~i~and~j~are~connected \\
 0 & otherwise.
\end{array}
\end{eqnarray}

A fully connected simple graph $K_V$ is a graph where each vertex is connected by
a single bond to any other vertex (beside itself):
$C_{i,j}=1-\delta_{i,j},~i,j\in G.$

A \emph{periodic orbit} (PO) on a graph G is a sequence of vertices, $[v_1,v_2,...,v_n]$
with $C_{v_i,v_{i+1}}=1$ and $v_1=v_n.$ PO's that can be obtained from one another by a cyclic permutation
of their vertices will be considered identical.

Consider an integer set $\{q_b\}_{b =(i,j)=(j,i)},~i,j \in G$ with $\sum q_b=n.$
A \emph{degeneracy class of $n$-bond periodic orbits} is a set of all the n-bond PO's each of which passes
exactly $q_b$ times over the bond $b.$
All these PO's are of the same length and since the bond lengths are rationally independent -
all PO's of the same length belong to one class.
The \emph{degeneracy} of a class is the number of distinct PO's in it.
Fig. 1.a shows the fully connected graph $G=K_5.$
Fig. 1.b shows the degeneracy class $(q_{(1,2)}=q_{(2,4)}=q_{(4,5)}=q_{(5,1)}=1, q_{(1,4)}=2).$
Fig. 1.c lists all the  PO's in this class and Fig 1.d shows two of them explicitly.

\begin{figure}
     \begin{center}
         \includegraphics[height=4.1in,width=3.2in,angle=270]{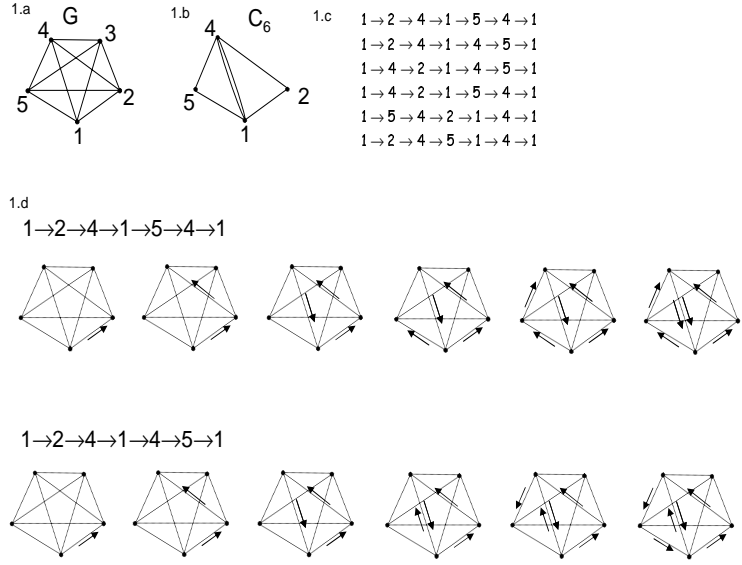}
     \end{center}
     \caption{The fully connected graph $K_5$ and one of its degeneracy class}
\end{figure}

Let $N_c(n,G)$ be the number of classes of $n$-bond
PO's in $G$ and $N_p(n,G)$ the total number of $n$-bond PO's in $G.$
The \emph{mean degeneracy} of n-bond PO in $G,$
$D_{n}(G),$ is defined by:
\begin{eqnarray}\label{1}
D_{n}(G)\equiv\frac{N_p(n,G)}{N_c(n,G)}.
\end{eqnarray}

  In the next section we shall provide exact expressions for $N_c(n,K_V),$ $N_p(n,K_V),$ and $D_{n}(K_V),$
that is, for the number of classes and PO's, and the mean degeneracy of
fully connected simple graphs.

\section{The number of classes of $n$-bond periodic
orbits on a fully connected graph with $V$ vertices }
Since a fully-connected graph is determined uniquely by specifying $V,$
we may use more compact notations, replacing $D_{n}(K_V)\rightarrow D(n,V)$ and so on.
We start by obtaining $N_c(n,V).$

Let $v\leq V$ and $N_{c,v}(n,v)$ be the number of $n$-bond degeneracy classes in $K_v$
which contain PO's that passes through all the $v$ vertices. Note that $N_{c,v}(n,v)\leq N_c(n,v)$ because not all
PO's  pass through all the vertices.
More precisely,
\begin{eqnarray}\label{2}
N_c(n,V)=\sum_{v=1}^V {V \choose v}N_{c,v}(n,v)
\end{eqnarray}
that is,  the number of classes is a sum of the number of classes
with PO's that use exactly $v$ vertices. The factor ${V \choose v}$
accounts for the possibilities of choosing these $v$ vertices.
All such choices have identical contribution since $K_V$ is fully connected.

For example consider $n=V=4.$ Figs.2.a, 2.b and 2.c show all the 4-bond
classes of $K_4$ grouped according to the sum in Eq.(\ref{2}).
There are ${4 \choose 2}=6$ ways for choosing 2 vertices (see
Fig.2.a), ${4 \choose 3}=4$ for choosing 3 vertices (corresponding
to each line in Fig.2.b), ${4 \choose 4}=1$ ways for choosing 4
vertices (Fig.2.c). These figures shows that, $N_{c,1}(4,1)=0,$
$N_{c,2}(4,2)=1,$ $N_{c,3}(4,3)=3,$ $N_{c,4}(4,4)=4.$
\begin{figure}
    \begin{center}
         \includegraphics[height=3.6in,width=3.75in,angle=270]{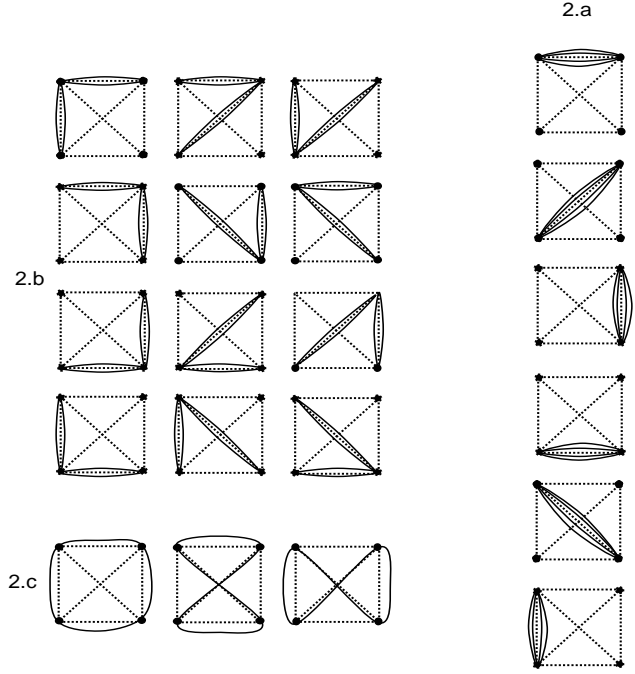}
    \end{center}
     \caption{4-bond degeneracy classes of the fully connected graph $K_4$}
\end{figure}

The reason why we have expressed $N_c(n,V)$ in Eq.(\ref{2}) in
terms of $N_{c,v}(n,v)$ is because the latter was calculated by R.
C. Read in Ref. \cite{Read}. For completeness, a concise
derivation along the lines of that work is given in Appendix 1.
The result derived in Appendix 1 is that $N_{c,v}(n,v)$ is $v!$
times the coefficient of $x^vt^n$ in the Taylor expansion (near
$t=x=0$) of $\ln \left( E(x,t)\right)$,
\begin{eqnarray}\label{3a}
N_{c,v}(n,v)=\frac{1}{n!}\frac{\partial^v}
{\partial x^v}\frac{\partial^n}{\partial t^n}\ln \left( E(x,t)\right)|_{x=0,t=0},
\end{eqnarray}
where
\begin{eqnarray}\label{3}
E(x,t)=\sum_{v=0}^{\infty}2^{-v}\frac{x^v}{v!}(1-t)^{-\frac{1}{2}v(v-1)}
\sum_{s=0}^{v}{v\choose s}\left(\frac{1-t}{1+t}\right)^{s(v-s)}.
\end{eqnarray}
From this and Eq.(\ref{2}) one sees that the number of classes of
$n$-bond periodic  orbits on a fully connected graph with $V$
vertices is
\begin{eqnarray}\label{4} N_c(n,V)=\frac{1}{n!}\sum_{v=1}^V {V \choose v}
\frac{\partial^v}{\partial x^v }\frac{\partial^n}{\partial t^n} \ln \left(E(x,t)\right)|_{x=0,t=0}.
\end{eqnarray}
An explicit expansion of $E(x,t)$ yields
\begin{eqnarray} \label{3b}
&&E(x,t)=\sum_{n,v=0..\infty }E_{n,v}t^nx^v
\end{eqnarray}
with
\begin{eqnarray}\label{3c}
&&E_{n,v}= \sum_{s=0..v}\frac{(-1)^{\mu}}{2^{v}v!}{ v\choose s} { \mu+s(v-s)-1\choose \mu} { n-\mu+{ s\choose 2}
 +{ v-s\choose 2} -1\choose  n-\mu}\nonumber  \\
 && ~~~~~~~~~~_{\mu=0..n}
\end{eqnarray}
One can also derive the following recursion relation (see Appendix 2)
\begin{eqnarray}\label{5recur}
N_{c,v}(n,v)=v!E_{n,v} -\sum_{m=0..n}\sum_{k=1..v-1} \frac{(v-1)!}{(k-1)!} N_{c,v}(m,k)E_{n-m,v-k}
\end{eqnarray}
 which enables a fast numerical calculation of $N_c(n,V).$

 As an example, consider the case $V=n=4.$
Substituting these values in Eq.(\ref{4}) one obtains
$N_c(4,4)=21.$ This result is confirmed in Figs.2.a-c which
present all the 21 classes of PO's.
Eq. A.1, in Appendix 1 of Ref. \cite{BK00} provides the asymptotic behavior, for large $n,$
 of the number of classes $N_c$ (for \emph{any} connected graph):
 \begin{eqnarray}
 \label{a18}
 N_c(2n,V)+N_c(2n+1,V)\sim \frac{2^{B-V+1}n^{B-1}}{(B-1)!}(1+O(\frac{1}{n})),
 \end{eqnarray}
 where $B$ is the number of bonds.
 for the complete-graph one has $B=V(V-1)/2.$ By substituting this value into Eq.(\ref{a18}) we verified numerically that
 the asymptotic   behavior of Eq.(\ref{4}) for $n\gg V$ matches Eq.(\ref{a18}).
 The results are shown in  Figure 3 which presents the ratio
 $(N_c(n,V)+N_c(n+1,V))/(N_c^{asymp}(n,V)+N_c^{asymp}(n+1,V)),$ for   even $n$.
 The superscript $asympt$ stands for the values obtained using Eq.(\ref{a18}).
 \begin{figure}
     \begin{center}
         \includegraphics[height=3.8in,width=3.0in,angle=270]{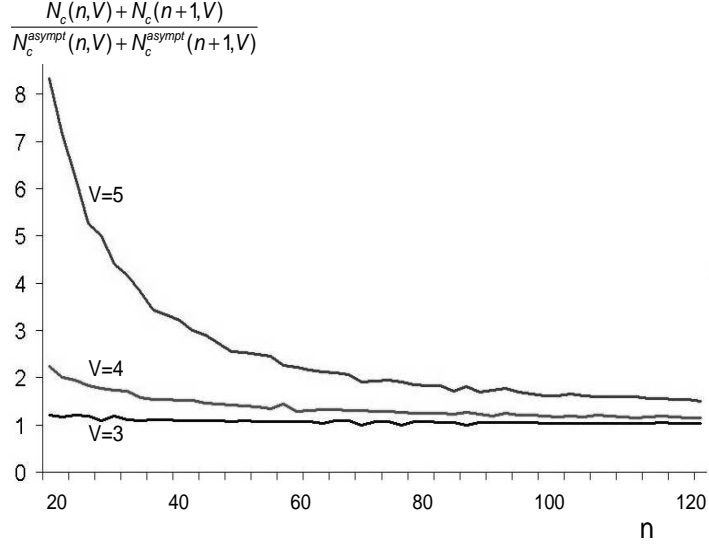}
   \end{center}
   \caption{Asymptotic behavior of the number of degeneracy classes}
 \end{figure}

\section{The mean degeneracy of $n$-bond periodic orbits
on a fully connected graph with $V$ vertices. Numerical results. }
To obtain the mean degeneracy Eq. (\ref{1}), we need also  to derive an expression
for $N_p(n,V)$ i.e. the number of $n$-bond PO's in $K_V.$ Let
us first calculate the number of $n$-bond closed trajectories.
The number of $n$-bond closed trajectories is given by
\begin{eqnarray}
\label{6} N(n,V)=TrC^n
\end{eqnarray}
where $C$ is the connectivity matrix defined in Eq.(\ref{connct}).
In our case it is given by:
\begin{eqnarray}\label{12}
C_{i,j}(K_V)=1-\delta_{ij},
\end{eqnarray}
and its eigenvalues are: $\lambda_1=V-1,~\lambda_2=\lambda_3=...\lambda_V=-1.$
From Eqs.(\ref{6}) and (\ref{12}) it follows that
\begin{eqnarray}\label{15}
N(n,V)=(V-1)^n +(V-1)(-1)^n.
\end{eqnarray}
$N(n,V)$ is the number of $n$-bond PO's but is different from
$N_p(n,V)$ in Eq. (\ref{1}) since in the latter, PO's that can be
obtained from one another through a cyclic permutation are
considered to be the same PO.

For simplicity, let us assume that $n$ is a prime number, thus avoiding
 the complications arising from the presence of PO's
which are repetitions of a shorter PO. With this assumption, each of the PO's counted in
Eq. (\ref{15}) is one of $n$ PO's that can be obtained from one
another by cyclic permutations of the vertices. Since we regard
all such cyclicly-equivalent PO's to be the same one, $N_p(n,V)$ and
$N(n,V)$ are related by
\begin{eqnarray}\label{16} N_p(n,V)=\frac{1}{n}N(n,V).
\end{eqnarray}

From Eqs. (\ref{1}) and (\ref{16}) one has
\begin{eqnarray}\label{17}
D(n,V)=\frac{(n-1)!((V-1)^n+(-1)^n(V-1))}
{\sum_{v=1}^V {V \choose v}\frac{\partial^v}
{\partial x^v}\frac{\partial^n}{\partial t^n}
\ln \left(E(x,t)\right)|_{x=0,t=0}}
\end{eqnarray}
  The mean-degeneracy $D(n,V)$ and its logarithm are shown in Figures 4, 5 and 6. These plots where
 generated using either Eq.(\ref{17}) or the recursive relation Eq.(\ref{5recur}).
Figure 4 presents the $n$-dependence of $D(n,V)$ for fixed values of $V.$
Figure 5 shows the $V$-dependence of $D(n,V)$ for fixed values of $n.$
As seen from these figures, the mean degeneracy grows rapidly and achieves values much larger than 2
already for small (i.e., much smaller than the number of bonds)  values of $n.$ On the other hand, in the limit
$V/n\rightarrow \infty $ it approaches 2  (most classes contain only a single PO and its time reversal).
Approximating
$ D(n,V)\approx \frac{N(n,V)/n + N(n,V)/(n+1)}{N_c(2n,V)+N_c(2n+1,V)} $
and then using the asymptotic expression Eq.(\ref{a18}) together with Eq.(\ref{15}) one has
\begin{eqnarray}
\label{a17}
D(n,V)\approx V(V^2-V-1)!2^{V-1}\frac{(V-1)^n}{n^{V(V-1)/2}}
\end{eqnarray}
Taking the logarithm of both sides and keeping only terms containing $n$ (assuming $\log n \ll V$)
one obtains
\begin{eqnarray}
\label{b17}
\log(D(n,V))\approx n \log(V-1)-\frac{1}{2}V(V-1)\log n.
\end{eqnarray}
taking the derivative with respect to $V$ yields the approximate value of $V$ in which
the maximal mean-degeneracy is obtained:
\begin{eqnarray}
\label{c17}
V_{max}\approx \sqrt{\frac{n}{\log n}}.
\end{eqnarray}
Although this estimation was derived for large $n$ it shows reasonable agreement with
the peaks in Figure 6. For example, the two maxima marked with arrows, for $n=20$ and $n=30,$
are located in the vicinity of $V=3.9$ and $V=4.5$ respectively, in agreement with Eq.(\ref{c17}).

\begin{figure}
     \begin{center}
         \includegraphics[height=4.1in,width=3.4in,angle=270]{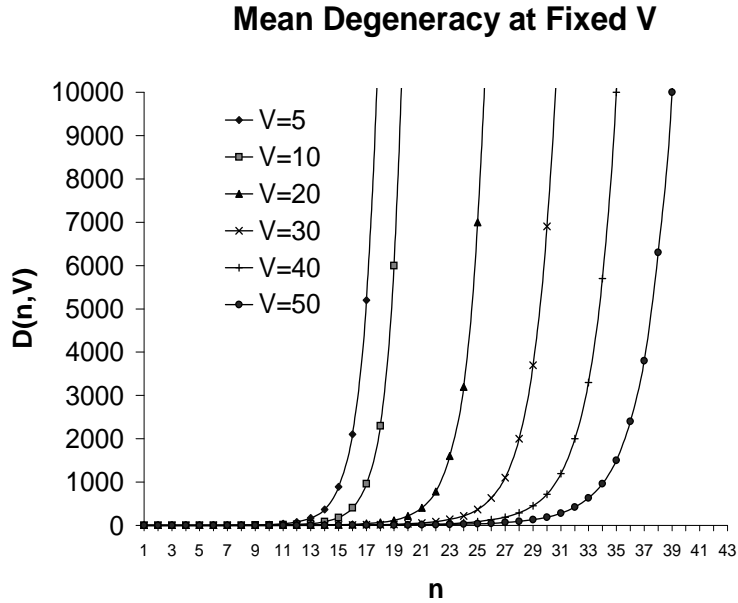}
   \end{center}
   \caption{The mean degeneracy for fixed number of vertices, $V.$}
 \end{figure}
 \begin{figure}
     \begin{center}
         \includegraphics[height=3.5in,width=2.8in,angle=270]{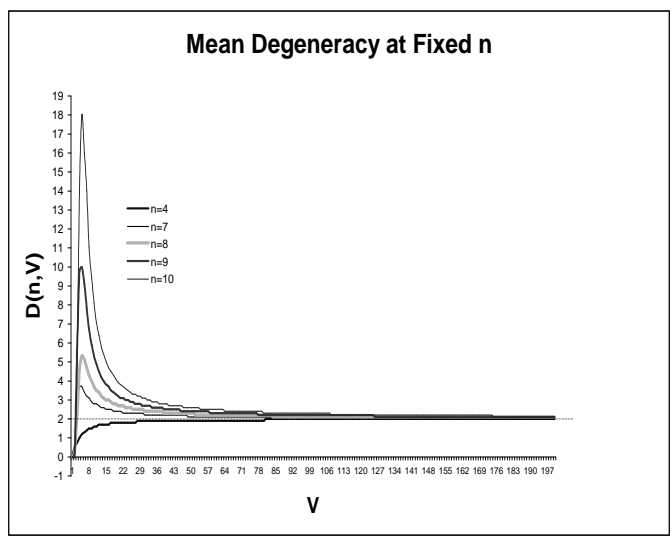}
   \end{center}
   \caption{The mean degeneracy at fixed PO's lengths.}
 \end{figure}
 \begin{figure}
     \begin{center}
         \includegraphics[height=3.5in,width=2.8in,angle=270]{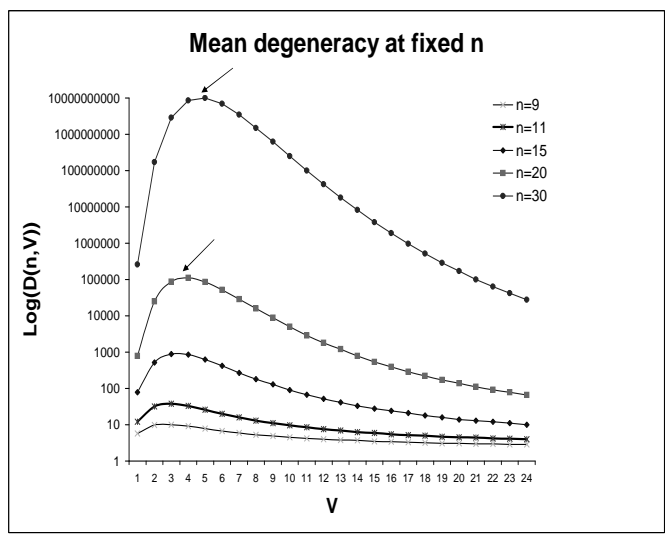}
   \end{center}
   \caption{The logarithm of the mean degeneracy at fixed PO's lengths.}
 \end{figure}

\subsection*{Acknowledgment}
We would like to thank G. Berkolaiko  for helpful discussions and K. Hornberger for help in the
implementation of the Maple code used for producing Fig. 5.
This work was supported in part by two Minerva Centers at the Weizmann Institute:
The Einstein Center for Theoretical Physics and The Center of Complex Systems,
and by the ANR grant number ANR-06-BLAN-0218-01.

\subsection*{Appendix 1. Derivation of Eq. (\ref{3a})}
In this appendix we derive Eq. (\ref{3a}) along  the lines of Ref. \cite{Read}.
\subsubsection*{Definitions}

A  \emph{graph} 
of order $V$ is a set of $V$ numbered vertices some of
which are connected by a bond 
(no more than one bond between two vertices, no bond connects a vertex to itself i.e. no loops).
The number of bonds connected to a vertex is the vertex \emph{valency.}Fig. 7.1 shows a graph of order 6.

\begin{figure}
    \begin{center}
         \includegraphics[height=5.2in,width=3.75in,angle=270]{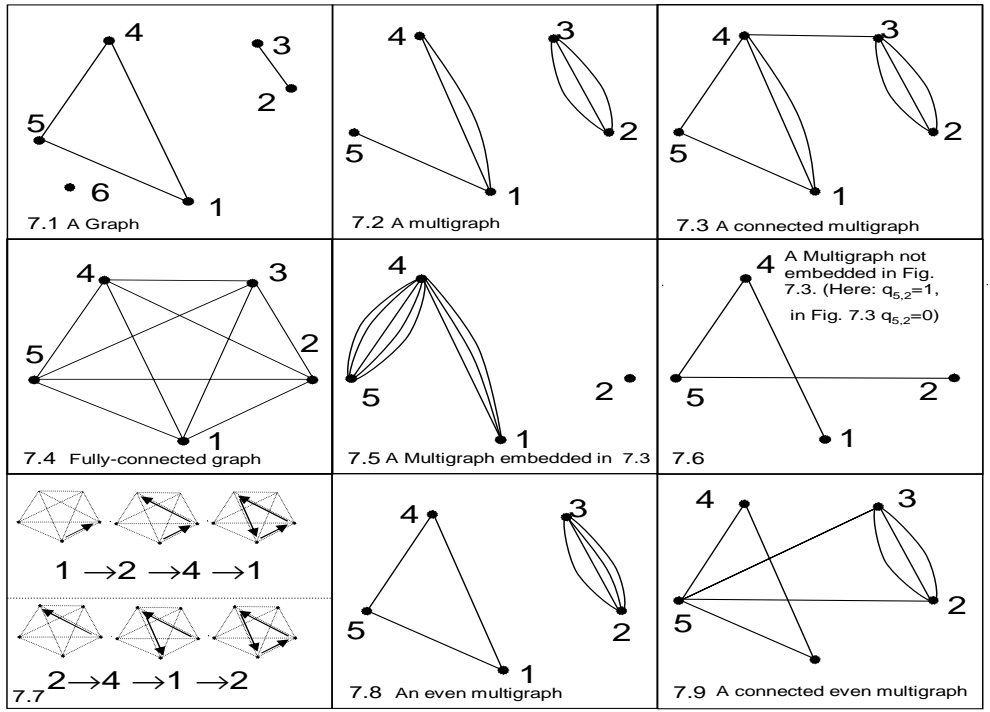}
     \end{center}
     \caption{Graphs and Multigraphs}
\vspace{-0.2cm}
\end{figure}

A \emph{multigraph} 
 is similar to a graph except  that there can
be more than one bond between two vertices (Fig. 7.2). A graph is a specific case of a multigraph.
Let $q_{(i,j)}(g)$ be the number of bonds connecting the vertices
$i$ and $j$ in a multigraph $g.$ We shall refer to $q_{(i,j)}(g)$ as
the \emph{degree of the bond $(i,j)$ in $g.$} For example, in
Fig. 7.2 $q_{(2,3)}=q_{(3,2)}=3.$ Note that the pairs (i,j) are \emph{not-directed}, 
that is, (i,j) and (j,i) are considered as the same bond.

A multigraph is \emph{connected} if, by moving on the bonds, one
can pass between any two of its vertices (in particular, all
valencies are $\geq 1,$ Fig. 7.3). A  graph, $K_V,$  is
\emph{fully-connected} if each of its $V$ vertices is connected to
all other vertices (Fig.7.4). Thus, $K_V$ has $V(V-1)/2$ bonds.

Let $g$ be a multigraph of order $v.$ $g$  is said  to be
\emph{embedded} in a graph $G$ if it can be obtained from $G$ by
first adding and deleting bonds between vertices \emph{that are
connected} in $G,$ and then  deleting some of the vertices which
have zero valency (now, \emph{after} the addition and deletion of
the bonds). By this definition, the order of $G$ is larger  or
equal to $v.$ If $q_{(i,j)}(G)=1$ then $q_{(i,j)}(g)=0,1,2...$ and if
$q_{(i,j)}(G)=0$ then $q_{(i,j)}(g)=0,$. The multigraph in Fig. 7.5
is embedded in the graph of Fig. 7.3 while the multigraph in
Fig. 7.6 is not.

A \emph{trajectory} 
is a sequence of vertices,
adjacent pairs of which are connected. If it is closed, i.e. it
starts and ends at the same vertex, the trajectory is a PO.
Actually, one can associate several PO's with each closed
trajectory since one can start in any of the trajectory points,
however, we shall refer to all of these as a single PO that is say
for example that $1\rightarrow 2\rightarrow 4\rightarrow 1$ is the
same PO as $ 2\rightarrow 4\rightarrow 1\rightarrow 2$ and two
PO's are distinct only if they can not be obtained from one
another by such a cyclic permutation. Thus, the two closed
trajectories in Fig. 7.7 are the same PO.

An \emph{even} multigraph is a multigraph where each vertex has an
even valency (Fig.7.8).  For any connected even multigraph (often called Euler multigraph) one
can always find a PO that passes on each bond exactly once (Eulerian circuit).
Often, there is more than one.
For example, the two PO's (Fig.1.d) $1\rightarrow
2\rightarrow 4\rightarrow 1\rightarrow 5\rightarrow 4\rightarrow
1$ and $1\rightarrow 2\rightarrow 4\rightarrow 1\rightarrow
4\rightarrow 5\rightarrow 1$ passes once on each bond in the connected even
multigraph $C_6$ shown in Fig. 1.b.

A \emph{class of $n$-bond periodic orbits}, $C_n,$ in a graph $G$
is an  $n$-bond connected even multigraph which is  embedded in a labeled
graph $G.$ $C_n,$ is specified by specifying the set  of
bond-degrees $\{q_{(i,j)}(C_n)\}_{i,j\in G},$ where $\sum_{i,j~j>i}
q_{(i,j)}(C_n)=n.$  A PO is said to be \emph{in} $C_n$ if it
consists of $n$ steps passing exactly $q_{(i,j)}(C_n)$ times between
$i$ and $j.$ By this definition, all PO's in $C_n$ have exactly the
same length, independently of the choice of bond lengths and
therefore, at a given energy, have the same action. Fig.1.b shows
a class of 6-bond PO's, $C_6,$ embedded in the fully-connected
graph $G$ in Fig.1.a. The PO's in this class are listed in
Fig.1.c. Two of them are drawn in Figs.1.d.

\subsubsection*{Proof of Eq. (\ref{3a})}

The proof of Eq. (\ref{3a}) is based on that given in Ref. \cite{Read}
which can also be used to treat the case of classes of graphs
with and without loops, and multigraphs with loops. Consider a set
of $n$ labeled vertices, and the set $\Omega(n,v)=\{G_{n,v}  \}$
 of multigraphs with $n$ bonds that one can draw on these
$v$ vertices (\emph{on} means using all of them  - i.e. these
multigraphs are of order $v$). To each of the $v$ vertices we
assign a sign, +1 or -1. There are $2^v$ such possible
assignments. For a given assignment $S$, we define the sign of
each bond in $G_{n,v}$ to be the product of signs of its two
vertices. The sign, $\sigma(G_{n,v},S),$ of a multigraph
$G_{n,v}\in \Omega(n,v)$   is then defined as the product of signs
of all its bonds. Thus,
\begin{eqnarray}
\label{App 1}\sigma(G_{n,v},S)=(-1)^{V_-(G_{n,v},S)}=(-1)^{\mu(G_{n,v},S)}
\end{eqnarray}
where $V_-$ is the sum of valencies of the negative vertices
and $\mu$ the number of negative bonds.
The sum of the signs of $G_{n,v}$ for all possible $S$ is $\sum_S (-1)^{V_-(G_{n,v})}.$
Summing this over all members of $\Omega(n,v)$
one has
\begin{eqnarray}\label{App 2}
\sum_{G_{n,v}\in\Omega(n,v)}
\left(\sum_S (-1)^{V_-(G_{n,v},S)}\right)=
\sum_S\left(\sum_{G_{n,v}\in\Omega(n,v)}(-1)^{\mu((G_{n,v},S))}\right).
\end{eqnarray}
In the right hand side the order of summation  was reversed and
Eq. (\ref{App 1}) was used. Consider the left hand side of Eq.
(\ref{App 2}). If $G_{n,v}$ is an even multigraph, then
$V_-(G_{n,v},S)$ is an even number for any $S$ and therefore
$\sum_S (-1)^{V_-(G_{n,v},S)}=2^v.$ If $G_{n,v}$ is not even, then
at least one of its vertices, say $A,$ has an odd valency. Since
for each assignment $S$ in which $A$ is negative there exists $S'$
which is identical to $S$ except that $A$ is positive in it, and
since $\sigma(G_{n,v},S)=-\sigma(G_{n,v},S'),$  one has $\sum_S
(-1)^{V_-(G_{n,v},S)}=0$ for any $G_{n,v}$ which is not even.
Thus, the left hand side of Eq. (\ref{App 2}) is the number of
even multigraphs in $\Omega(n,v)$ times $2^v.$ To obtain the
right-hand side, consider the ${v\choose s}$ assignments in which
exactly $s$ of the vertices are positive. The number of ways to
put $\mu$ identical balls in $s(v-s)$ identical boxes each of
which may contain any number of balls, is   ${\mu+s(v-s)-1\choose
\mu}$ and therefore this is the number of ways the
$\mu(G_{n,v},S)$ bonds which join the $s$ positive with the $v-s$
negative vertices can be placed. The remaining $n-\mu$ bonds may
be placed between the ${s\choose 2}+{v- s\choose 2}$ pairs of
vertices with identical signs, that is in
\begin{eqnarray}
{n-\mu+{s\choose 2}+{v-s \choose 2}-1\choose n-\mu}
\end{eqnarray}
different ways. (To enable compact writing, here and below,  we
assume that  the binomial coefficients have the properties ${a
\choose b}=0$ for $b>a$ \emph{and} $b\neq 0,$ and ${a \choose
0}=1$ for any $a$). Summing over all possible $\mu(G_{n,v},S)$ one
gets the total contribution of all assignments in which exactly
$s$ vertices are positive:
\begin{eqnarray}\label{App 4}
\sum_{\mu=0}^n (-1)^{\mu}{\mu+s(v-s)-1\choose \mu}
{n-\mu+{s\choose 2}+{v-s \choose 2}-1\choose n-\mu}.
\end{eqnarray}
This contribution is the coefficient of $t^n$ in  $(1-t)^{(-1/2)v
(v-1)}   (\frac{1-t}{1+t})^{s(v-s)}.$ Thus, the number of $n$-bond
even multigraphs one can draw on $v$ labeled vertices is  given
by $v!$ times the coefficient of $t^nx^v$ in the power expansion
of:
\begin{eqnarray}
E(x,t)=\sum_{v=0}^{\infty}2^{-v}\frac{x^v}{v!}(1-t)^{-\frac{1}{2}v(v-1)}
\sum_{s=0}^{v}{v\choose s}\left(\frac{1-t}{1+t}\right)^{s(v-s)}.
\end{eqnarray}
We are interested in $N_{c,v}(n,v),$ i.e. the number  $n$-bond
\emph{connected} even multigraphs one can draw on $v$ labeled
vertices. It is a known result in graph enumeration theory that
the generating function of the connected set of (labeled) graphs
is given by the log of that of the non-connected set \cite{Harary}. Thus, $N_{c,v}(n,v),$ is $v!$ times the coefficient
of $t^nx^v$ in the power expansion of $\ln(E(x,t))$ which proves
Eq.(\ref{3a}).

\subsection*{Appendix 2. Proof of Eq.(\ref{5recur}) }
Define the expansions
\begin{eqnarray}\label{eq5b}
E(x,t)=\sum_{v=0}^{\infty}E_v(t)x^{v},
\end{eqnarray} and
\begin{eqnarray}\label{eq6}
\ln(E(x,t))=\sum_{v=1}^{\infty}L_v(t)\frac{x^{v}}{v!}.
\end{eqnarray} $N_{c,v}(n,v)$ is $v!$ times the coefficient
of $x^vt^n$ in $\ln(E(x,t))$ and therefore
\begin{eqnarray}\label{eq7}
L_v(t)=\sum_{n=0}^{\infty}N_{c,v}(n,v)t^n.
\end{eqnarray}
There exists a useful recursion relation between the coefficients in Eqs. ( \ref{eq5b}) and (\ref{eq6}):
\begin{eqnarray}\label{eq8}
&&L_v(t)=v!E_v(t)-\sum_{k=1}^{v-1}\frac{(v-1)!}{(k-1)!}L_k(t)E_{v-k}(t)~~v> 1\nonumber \\
&&L_1(t)=E_1(t)~~~
\end{eqnarray}
Expansion of Eqs.(\ref{eq6}) and (\ref{eq7}) in powers of $t$ yields Eq.(\ref{5recur}).

\end{document}